\documentclass[iop,onecolumn,numberedappendix,appendixfloats]{emulateapj}

\usepackage{amsmath,amssymb,bm,mathrsfs}

\newcommand{\fint}{\;{-}\kern-1.07em\intop\nolimits}
\newcommand{\tfint}{\mbox{${-}\kern-.9em\intop$}}

\begin{document}

\title{Laboratory Frequency Redistribution Function for\break
the Polarized $\Lambda$-Type Three-Term Atom}

\author{R.~Casini$^1$ and R.~Manso Sainz$^2$}

\affil{$^1$High Altitude Observatory, National Center for Atmospheric
Research,\footnote{The National Center for Atmospheric Research is sponsored
by the National Science Foundation.}\break
P.O.~Box 3000, Boulder, CO 80307-3000, U.S.A.}
\affil{$^2$Max Planck Institute for Solar System Research,\break
Justus-von-Liebig-Weg 3, 37077 G\"ottingen, Germany}

\begin{abstract}
We present the frequency redistribution function for the 
polarized three-term atom of the $\Lambda$-type in the collisionless 
regime, and we specialize it to the case where both the initial 
and final terms of the three-state transition are metastable (i.e., 
with infinitely sharp levels). This redistribution function represents 
a generalization of the well-known $R_{\rm II}$ function to the case 
where the lower terms of the transition can be polarized and carry 
atomic coherence, and it can be applied to the investigation of 
polarized line formation in tenuous plasmas, where collisional rates 
may be low enough that anisotropy induced atomic polarization 
survives even in the case of metastable levels.
\end{abstract}

\section{Introduction}

Plasma diagnostics in both laboratory and astrophysical settings relies
principally on spectroscopic and polarimetric methods of observation, where
the radiation emitted from the plasma is analyzed as a function of
wavelength and state of polarization. These allow one to determine important
properties of the observed plasma, such as bulk and turbulent velocity 
fields, gas composition and density, existence of anisotropic processes
such as the presence of directed electric and/or magnetic fields, and
anisotropic sources of irradiation and of colliding particles.

Particularly in the case of low density gases, collisional processes can
be of such modest magnitude that the excitation state of the plasma is
typically far away from local thermodynamical equilibrium (LTE). In that
case, the excitation and de-excitation processes that are responsible
for the observed radiation may be statistically correlated, so that the
system is able to preserve a ``memory'' of the excitation conditions. 
A typical example is the partially coherent scattering of 
radiation, where the frequency of the outgoing photon is correlated
(via an intrinsically second-order atom--photon process) to the spectral 
distribution of the incoming radiation, leading to the phenomenon
of partial redistribution of frequency.

These excitation conditions are rather common in low-density astrophysical 
plasmas, such as the higher layers of the solar atmosphere, planetary 
nebulae, and interstellar \ion{H}{1} regions, where outstanding spectral 
lines of the observed spectrum (e.g., the first few lines of the 
Lyman and Balmer series of hydrogen, some lines of neutral sodium, and 
of singly ionized calcium, etc.) are known to be formed in
conditions of strong departure from LTE, which may often also include 
the effects of partially coherent scattering of radiation.

\begin{figure}[t!]
\centering
\includegraphics[width=.45\hsize]{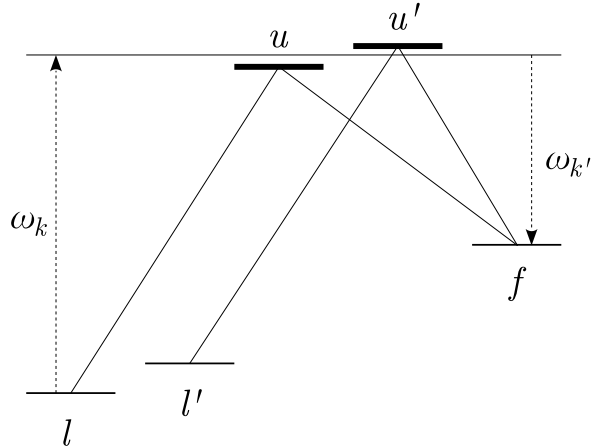}
\caption{Schematic diagram for the fluorescent scattering in a 
three-term model atom of the $\Lambda$-type, for an incoming 
photon of frequency $\omega_k$ and an outgoing photon of frequency
$\omega_{k'}$. The model atom considered in this paper is restricted 
to the case where all $(l,l')$ and $f$ levels are sharp (i.e., 
metastable). However, all the levels involved can be non-degenerate.
\label{fig:Lambda-model}}
\end{figure}

In this paper, we derive the redistribution function in the 
laboratory frame for the polarized three-term atom of the $\Lambda$-type, 
i.e., for the transition system $(l,l')\to (u,u')\to f$ such that
$l,l',f\prec u,u'$ by energy ordering (see 
Figure~\ref{fig:Lambda-model}). This model can be used to investigate, 
for example, the formation of the \ion{Ca}{2} system of transitions 
comprising the K and H lines and the infrared (IR) triplet of the solar spectrum 
\citep{CM16}.

The expression we arrive at corresponds to the extension of the
well-known $R_{\rm II}$ redistribution function \cite[e.g.,][]{Mi78}
to the case of a three-term atom that can harbor atomic polarization in
all of its levels, including those of the lower (metastable) states.
In applications where plasma collisions are important in determining 
the statistical equilibrium of the atomic system, this expression is 
still useful as it provides the correct contribution to the total 
emissivity of the plasma from radiation scattering. As long as the 
collisional lifetime of the metastable states is much longer than 
the \emph{total} lifetime of the upper state, the approximation of 
sharp lower levels remains applicable in practice, and our expression 
of $R_{\rm II}$ can then be used, with proper weights, alongside with 
$R_{\rm III}$, to completely describe the radiative and collisional 
redistribution of radiation in the modeled atmosphere 
\cite[see, e.g.,][]{BT12}.

\section{The redistribution function for the $\Lambda$-type three-term atom}
\label{sec:derivation}

\renewcommand*{\thefootnote}{\arabic{footnote}}
\setcounter{footnote}{0}

We rely on previously published work \cite[][hereafter, Paper~I]{Ca14} 
on the redistribution function for the polarized two-term atom
in order to develop the framework for the generalization of
$R_{\rm II}$ to the case of a three-term atom of the 
$\Lambda$-type, undergoing the two-photon transition 
$(l,l')\to (u,u')\to f$ (see Figure~\ref{fig:Lambda-model}). The
bases for this generalization were laid out in a recent paper 
\citep{CM16}, where we showed how the radiative transfer equation 
describing the polarized line formation in a two-term atom in the
presence of partially coherent scattering (Equations (19) and (20) 
of Paper~I) naturally extends to the case of the multi-term 
atom of the $\Lambda$-type (Equations (1) and (2) of
\citealt{CM16}).

In practice, the main complication in deriving 
the redistribution function for the three-term atom comes from
allowing $\epsilon_f\ne\epsilon_{l,l'}$ for the final state $f$
(here $\epsilon_a$ is the total width of the level $a$ 
due to all possible relaxation processes), and from the need to 
assume generally different thermal widths for the two 
transitions $l\to u$ and $u\to f$.
Accordingly, we introduce a characteristic Doppler width $\Delta_{mn}$ 
for the atomic transition between two generic terms $m$ and $n$ with
energy $E_m$ and $E_n$, respectively.
We indicate with $\omega_{mn}=(E_m-E_n)/\hbar$ the Bohr frequency 
of such transition.

For each scattering event, the angle $\Theta$ between the propagation 
directions of the incoming and outgoing photons is an essential 
parameter of the laboratory frame redistribution function.
We introduce the associated quantities
\begin{equation}
C=\cos\Theta\;,\qquad S=\sin\Theta\;,
\end{equation}
and follow \cite{Hu82} for the choice of the Cartesian reference frame
for the projection of the thermal velocity of the plasma. This coincides
with the frame adopted in Paper~I \cite[see also][]{Mi78} when the 
thermal widths of the two transitions are identical (i.e.,
when the terms $l$ and $f$ have the same energy).
Accordingly, we define for the $\Lambda$-type transition $l\to u\to f$,
\begin{equation} \label{eq:avg_Doppler}
\Delta=(\Delta_{ul}^2+\Delta_{uf}^2 - 2 C \Delta_{ul}\Delta_{uf})^{1/2}\;,
\end{equation}
\begin{equation} \label{eq:xi}
\xi_l=\Delta_{ul}/\Delta\;,\qquad
\xi_f=\Delta_{uf}/\Delta\;.
\end{equation}
We note that Equations (\ref{eq:avg_Doppler}) and (\ref{eq:xi}) imply
\begin{equation} \label{eq:cool}
\xi_l^2+\xi_f^2 - 2 C \xi_l\xi_f=1\;.
\end{equation}

We define next the complex profile function \cite[e.g.,][]{LL04}
\begin{eqnarray} \label{eq:W}
W(v,a)&\equiv&
\frac{1}{\pi}\int_{-\infty}^{+\infty} dp\;
\frac{{\rm e}^{-p^2}}{a+{\rm i}(p-v)}
= \exp(a-{\rm i}v)^2\,\mbox{erfc}(a-{\rm i}v) \nonumber \\
&\equiv& H(v, a)+{\rm i}\,L(v, a)\;,
\end{eqnarray}
where $\mbox{erfc(z)}$ is the complementary error function 
\cite[e.g.,][]{AS64},
and $H(v,a)$ and $L(v,a)$ are respectively the Voigt and Faraday--Voigt 
functions, along with the dimensionless frequency variables
\begin{equation} \label{eq:variables.1}
v_{mn}=(\hat\omega_k-\omega_{mn})/\Delta\;,\qquad
w_{mn}=(\hat\omega_{k'}-\omega_{mn})/\Delta\;.
\end{equation}
The incoming and outgoing radiation frequencies, $\hat\omega_k$ and 
$\hat\omega_{k'}$, are expressed in the laboratory frame of reference, 
and they differ from the corresponding frequencies in the atomic frame of 
rest because of the thermal motion of the atoms (see Appendix of Paper~I). 
We also introduce normalized damping parameters associated with the inverse
lifetimes of the transition levels, using the same ``reduced'' Doppler width
of Equation (\ref{eq:avg_Doppler})
\begin{equation}
\label{eq:variables.2}
a_m=\epsilon_m/\Delta\;.
\end{equation}
In the presence of pressure broadening, the numerator of 
Equation (\ref{eq:variables.2}) should be augmented by the corresponding 
collisional inverse lifetime.

We note in passing that the proposed normalizations (\ref{eq:variables.1})
and (\ref{eq:variables.2}) differ from the one adopted in Paper~I, where 
instead the frequency and inverse lifetimes pertaining to a
given transition were normalized to the Doppler width for that
transition. The choice adopted 
here is dictated exclusively by convenience, as it significantly simplifies 
the form of the resulting expressions.

Following
the approach of Paper~I, Appendix~A, we find that we can perform the 
integration over two of the three components of the thermal velocity in 
the adopted reference frame. Accordingly, after some algebra, 
the redistribution function of the polarized three-term atom of the 
$\Lambda$-type can be written in the following integral form 
\begin{eqnarray} \label{eq:rlab}
R(\Omega_u,\Omega_{u'};\Omega_l,\Omega_{l'},\Omega_{f};
	\hat{\omega}_{k},\hat{\omega}_{k'};\Theta) &=&
\frac{1}{\Delta^2\,S \xi_l \xi_f} 
\int_{-\infty}^{+\infty} dq\; {\rm e}^{-q^2} \nonumber \\
&&\kern -2.2in\times\Biggl\{
\frac{W\bigl( \frac{v_{ul'}+q(C\xi_l\xi_f-\xi_l^2)}{S\xi_l\xi_f},
	\frac{a_u+a_{l'}}{S\xi_l\xi_f} \bigr)}
{a_{l'}+a_f+{\rm i}\bigl(q-v_{ul'}+w_{uf}\bigr)}
+
\frac{\overline{W}\bigl( \frac{v_{u'l}+q(C\xi_l\xi_f-\xi_l^2)}{S\xi_l\xi_f},
	\frac{a_{u'}+a_l}{S\xi_l\xi_f}\bigr)}
{a_l+a_f-{\rm i}\bigl(q-v_{u'l}+w_{u'f}\bigr)} 
	\nonumber \\
&&\kern -2.2in \,\,{}+
\frac{W\bigl( \frac{w_{uf}+q(\xi_f^2-C\xi_l\xi_f)}{S\xi_l\xi_f},
	\frac{a_u+a_f}{S\xi_l\xi_f} \bigr)}
{a_l+a_f-{\rm i}\bigl(q-v_{ul}+w_{uf}\bigr)}
+
\frac{\overline{W}\bigl( \frac{w_{u'f}+q(\xi_f^2-C\xi_l\xi_f)}{S\xi_l\xi_f},
	\frac{a_{u'}+a_f}{S\xi_l\xi_f}\bigr)}
{a_{l'}+a_f+{\rm i}\bigl(q-v_{u'l'}+w_{u'f}\bigr)} 
	\nonumber \\
&&\kern -2.2in\,\,{} +
\frac{W\bigl( \frac{v_{ul'}+q(C\xi_l\xi_f-\xi_l^2)}{S\xi_l\xi_f},
	\frac{a_u+a_{l'}}{S\xi_l\xi_f} \bigr)
    - W\bigl( \frac{w_{uf}+q(\xi_f^2-C\xi_l\xi_f)}{S\xi_l\xi_f},
	\frac{a_u+a_f}{S\xi_l\xi_f} \bigr)}
{a_f-a_{l'} - {\rm i}(q-v_{ul'}+w_{uf})} \nonumber \\
&&\kern -2.2in \,\,{}+
\frac{\overline{W}\bigl( \frac{v_{u'l}+q(C\xi_l\xi_f-\xi_l^2)}{S\xi_l\xi_f},
	\frac{a_{u'}+a_l}{S\xi_l\xi_f} \bigr)
     -\overline{W}\bigl( \frac{w_{u'f}+q(\xi_f^2-C\xi_l\xi_f)}{S\xi_l\xi_f},
	\frac{a_{u'}+a_f}{S\xi_l\xi_f} \bigr)}
{a_f-a_l + {\rm i}(q-v_{u'l}+w_{u'f})}
\Biggr\}\;,
\end{eqnarray}
where we have indicated with $\overline{W}$ the complex conjugate of $W$.
Equation (\ref{eq:rlab}) transforms exactly into Equation (A6) of 
Paper~I when $\xi_l=\xi_f$ (and letting $a_u=a_{u'}$ and
$a_l=a_{l'}=a_f$), noting that Equation (\ref{eq:avg_Doppler}) 
gives $\Delta=2 S_2\Delta\omega_T$, in such case, 
using the notation of Paper~I.

In order to facilitate further manipulation of Equation (\ref{eq:rlab}) for 
specific applications, it is convenient to introduce barred quantities 
${\bar x}=x/(S\xi_l\xi_f)$. Thus Equation (\ref{eq:rlab}) becomes
\begin{eqnarray*}
R(\Omega_u,\Omega_{u'};\Omega_l,\Omega_{l'},\Omega_{f};
	\hat{\omega}_{k},\hat{\omega}_{k'};\Theta) &=&
\frac{1}{\Delta^2\,S\xi_l\xi_f} 
\int_{-\infty}^{+\infty} d\bar q\; {\rm e}^{-S^2\xi_l^2\xi_f^2\bar q^2} \\
&&\kern -2.2in\times\Biggl\{
\frac{W\bigl( \bar v_{ul'}+\bar q(C\xi_l\xi_f-\xi_l^2),
	\bar a_u+\bar a_{l'} \bigr)}
{\bar a_{l'}+\bar a_f+{\rm i}\bigl(\bar q
	-\bar v_{ul'}+\bar w_{uf}\bigr)}
+
\frac{\overline{W}\bigl( \bar v_{u'l}+\bar q(C\xi_l\xi_f-\xi_l^2),
	\bar a_{u'}+\bar a_l \bigr)}
{\bar a_l+\bar a_f-{\rm i}\bigl(\bar q
	-\bar v_{u'l}+\bar w_{u'f}\bigr)} 
	\nonumber \\
&&\kern -2.2in \,\,{}+
\frac{W\bigl( \bar w_{uf}+\bar q(\xi_f^2-C\xi_l\xi_f),
	\bar a_u+\bar a_f \bigr)}
{\bar a_l+\bar a_f-{\rm i}\bigl(\bar q
	-\bar v_{ul}+\bar w_{uf}\bigr)}
+
\frac{\overline{W}\bigl( \bar w_{u'f}+\bar q(\xi_f^2-C\xi_l\xi_f),
	\bar a_{u'}+\bar a_f \bigr)}
{\bar a_{l'}+\bar a_f+{\rm i}\bigl(\bar q
	-\bar v_{u'l'}+\bar w_{u'f}\bigr)} 
	\nonumber \\
&&\kern -2.2in\,\,{} +
\frac{W\bigl( \bar v_{ul'}+\bar q(C\xi_l\xi_f-\xi_l^2),
	\bar a_u+\bar a_{l'} \bigr)
    - W\bigl( \bar w_{uf}+\bar q(\xi_f^2-C\xi_l\xi_f),
	\bar a_u+\bar a_f \bigr)}
{\bar a_f-\bar a_{l'} - {\rm i}(\bar q
	-\bar v_{ul'}+\bar w_{uf})} \nonumber \\
&&\kern -2.2in \,\,{}+
\frac{\overline{W}\bigl( \bar v_{u'l}+\bar q(C\xi_l\xi_f-\xi_l^2),
	\bar a_{u'}+\bar a_l \bigr)
     -\overline{W}\bigl( \bar w_{u'f}+\bar q(\xi_f^2-C\xi_l\xi_f),
	\bar a_{u'}+\bar a_f \bigr)}
{\bar a_f-\bar a_l + {\rm i}(\bar q
	-\bar v_{u'l}+\bar w_{u'f})}
\Biggr\}\;.
\end{eqnarray*}
We then define
\begin{equation} \label{eq:x.y.def}
x_{ab}=\bar v_{ab}+\bar q(C\xi_l\xi_f-\xi_l^2)\;,\qquad
y_{ab}=\bar w_{ab}+\bar q(\xi_f^2-C\xi_l\xi_f)\;,
\end{equation}
from which (see Equation (\ref{eq:cool}))
\begin{equation} \label{eq:ganzo}
x_{ab}-y_{ac}=\bar v_{ab}-\bar w_{ac}-\bar q\;.
\end{equation}
When $C=\xi_l/\xi_f$ or $C=\xi_f/\xi_l$, the 
coefficient of $\bar q$ in one of the two definitions (\ref{eq:x.y.def}) 
vanishes.\footnote{Note that this can happen for \emph{both}
definitions at the same time only if $\xi_f=\xi_l$, but in that case 
the problem becomes identical to the one for the two-term atom \citep{Ca14}.} 
However, the relation (\ref{eq:ganzo}), on which we rely for
the following development, remains valid.
Using these two definitions, we then can rewrite at last
\begin{eqnarray} \label{eq:rlab.bar}
R(\Omega_u,\Omega_{u'};\Omega_l,\Omega_{l'},\Omega_{f};
	\hat{\omega}_{k},\hat{\omega}_{k'};\Theta) &=&
\frac{1}{\Delta^2\,S\xi_l\xi_f} 
\int_{-\infty}^{+\infty} d\bar q\; {\rm e}^{-S^2\xi_l^2\xi_f^2\bar q^2} 
	\nonumber \\
&&\kern -2.2in\times\Biggl\{
\frac{W\bigl( x_{ul'},\bar a_u+\bar a_{l'} \bigr)}
{\bar a_{l'} + \bar a_f - {\rm i}(x_{ul'}-y_{uf})}
+
\frac{\overline{W}\bigl( x_{u'l},\bar a_{u'}+\bar a_l \bigr)}
{\bar a_l + \bar a_f + {\rm i}(x_{u'l}- y_{u'f})}
+\frac{W\bigl( y_{uf},\bar a_u+\bar a_f \bigr)}
{\bar a_l + \bar a_f + {\rm i}(x_{ul}- y_{uf})}
+
\frac{\overline{W}\bigl( y_{u'f},\bar a_{u'}+\bar a_f \bigr)}
{\bar a_{l'} + \bar a_f - {\rm i}(x_{u'l'}- y_{u'f})}
\nonumber \\
&&\kern -2.2in \,\,{}+
\frac{W\bigl( x_{ul'},\bar a_u+\bar a_{l'} \bigr)
	- W\bigl( y_{uf},\bar a_u+\bar a_f \bigr)}
{\bar a_f - \bar a_{l'} + {\rm i}(x_{ul'}-y_{uf})}
+
\frac{\overline{W}\bigl( x_{u'l},\bar a_{u'}+\bar a_l \bigr)
	-\overline{W}\bigl( y_{u'f},\bar a_{u'}+\bar a_f \bigr)}
{\bar a_f - \bar a_l - {\rm i}(x_{u'l}-y_{u'f})}
\Biggr\}\;.
\end{eqnarray}
This is the starting point for the derivation of the redistribution
function for the model atom considered in this paper.

\section{The case of the $\Lambda$-type three-term atom with metastable
lower states}

In order to treat the case of a $\Lambda$-type atom undergoing the 
transition $(l,l')\to (u,u')\to f$, where the initial and final terms are
metastable, we must consider the limit $\bar a_{l,l',f}\to 0$ of
Equation (\ref{eq:rlab.bar}), using the identity
\begin{equation} \label{eq:zeta0}
\lim_{\epsilon\to0^+}\frac{1}{\epsilon\pm {\rm i} z} =
\pi\delta(z)\mp{\rm i}\,{\rm Pv}\frac{1}{z}\;,
\end{equation}
where $\delta(x)$ and $\mbox{Pv}$ are, respectively, the Dirac delta 
and the Cauchy principal value distributions.
In general, we will assume that the
initial lower term is polarized and carrying atomic coherence 
(i.e., $\rho_{ll'}\ne0$). After some algebraic manipulation
(see Appendix~\ref{app:A}), and using
\begin{equation} \label{eq:useful}
v_{ul}-w_{uf} 
=\frac{\hat\omega_k-\hat\omega_{k'}+\omega_{lf}}{\Delta}
=v_{u'l}-w_{u'f}\;,
\end{equation}
we obtain
\begin{eqnarray} \label{eq:R.last}
R(\Omega_u,\Omega_{u'};\Omega_l,\Omega_{l'},\Omega_{f};
	\hat{\omega}_{k},\hat{\omega}_{k'};\Theta)_{\rm s.l.l.} 
&=& \frac{\pi}{\Delta^2\,S\xi_l\xi_f} \nonumber \\
&&\kern -2.5in \times \Biggl\{
\exp\biggl[-\frac{(\hat\omega_k-\hat\omega_{k'}+\omega_{lf})^2}
	{\Delta^2}\biggr]\,\biggl[
W\biggl(\frac{\kappa^+ v_{ul}+\kappa^- w_{uf}}{S\xi_l\xi_f},
	\frac{a_u}{S\xi_l\xi_f} \biggr) +
\overline{W}\biggl(\frac{\kappa^+ v_{u'l}+\kappa^- w_{u'f}}{S\xi_l\xi_f},
	\frac{a_{u'}}{S\xi_l\xi_f} \biggr) \biggr] \nonumber \\
&&\kern -2.4in \mathop{+}
\exp\biggl[-\frac{(\hat\omega_k-\hat\omega_{k'}+\omega_{l'f})^2}
	{\Delta^2}\biggr]\,\biggl[
W\biggl(\frac{\kappa^+ v_{ul'}+\kappa^- w_{uf}}{S\xi_l\xi_f},
	\frac{a_u}{S\xi_l\xi_f} \biggr) +
\overline{W}\biggl(\frac{\kappa^+ v_{u'l'}+\kappa^- w_{u'f}}{S\xi_l\xi_f},
	\frac{a_{u'}}{S\xi_l\xi_f} \biggr) \biggr]
	\Biggr\} \nonumber \\
&&\kern -2.5in \,\mathop{+}
\frac{\rm i}{\Delta^2\,S\xi_l\xi_f} 
\fint_{-\infty}^{+\infty} dq\; {\rm e}^{-q^2}
\biggl(
\frac{1}{v_{ul'}-w_{uf}-q}-
\frac{1}{v_{ul}-w_{uf}-q} \biggr) \nonumber \\
	\nonumber \\
&&\kern -1.5in \times \biggl[
W\biggl( \frac{w_{uf}
	+q(\xi_f^2-C\xi_l\xi_f)}{S\xi_l\xi_f},
	\frac{a_u}{S\xi_l\xi_f} \biggr) +
\overline{W}\biggl( \frac{w_{u'f}
	+q(\xi_f^2-C\xi_l\xi_f)}{S\xi_l\xi_f},
	\frac{a_{u'}}{S\xi_l\xi_f} \biggr)
	\biggr]\;,
\end{eqnarray}
where we introduced the quantities defined in Equation (\ref{eq:chi}), and 
where the symbol $\tfint$ indicates that the integral must be evaluated
as the Cauchy principal value.
We note that it is possible to rewrite
\begin{eqnarray} \label{eq:alter}
&&\biggl(
\frac{1}{v_{ul'}-w_{uf}-q}-
\frac{1}{v_{ul}-w_{uf}-q} \biggr) 
\biggl[
W\biggl( \frac{w_{uf}+q(\xi_f^2-C\xi_l\xi_f)}{S\xi_l\xi_f},
	\frac{a_u}{S\xi_l\xi_f} \biggr) +
\overline{W}\biggl( \frac{w_{u'f}+q(\xi_f^2-C\xi_l\xi_f)}{S\xi_l\xi_f},
	\frac{a_{u'}}{S\xi_l\xi_f} \biggr)
	\biggr] \nonumber \\
&&\kern 2in \equiv
\frac{\omega_{ll'}}{\Delta}\,
\frac{
W\Bigl( \frac{w_{uf}+q(\xi_f^2-C\xi_l\xi_f)}{S\xi_l\xi_f},
	\frac{a_u}{S\xi_l\xi_f} \Bigr) +
\overline{W}\Bigl( \frac{w_{u'f}+q(\xi_f^2-C\xi_l\xi_f)}{S\xi_l\xi_f},
	\frac{a_{u'}}{S\xi_l\xi_f} \Bigr) }%
{(v_{ul'}-w_{uf}-q) (v_{ul}-w_{uf}-q)}\;,
\end{eqnarray}
where we observed that
\begin{displaymath}
v_{ul}-v_{ul'}
=\frac{\omega_{ll'}}{\Delta}
=v_{u'l}-v_{u'l'}\;.
\end{displaymath}
Thus, the integral term in Equation (\ref{eq:R.last}) 
vanishes when $\omega_{ll'}=0$, i.e., for completely degenerate 
lower levels, or in the case of non-coherent lower term (see
below).
In Appendix~\ref{app:C}, we show
that this integral contribution is simply a 
frequency redistribution term that carries no net energy.

Despite the relative simplicity of Equation (\ref{eq:alter}), the original 
form of the integrand as given in Equation (\ref{eq:R.last}) is more
convenient for numerical computation, since it can 
be shown that (see Appendix~\ref{app:B})
\begin{eqnarray}
&&\fint_{-\infty}^{+\infty} dq\;
\frac{{\rm e}^{-q^2}}{v-w-q}\,
W\biggl(\frac{w+q(\xi_f^2-C\xi_l\xi_f)}{S\xi_l\xi_f},
	\frac{a}{S\xi_l\xi_f}\biggr) \nonumber \\
&=&
\frac{2}{\sqrt{\pi}}
\int_{-\infty}^{+\infty} dp\;
\frac{{\rm e}^{-p^2/\xi_f^2}}{a+{\rm i}(p-w)}\,
F\biggl(\frac{v-w+p(1-C\xi_l/\xi_f)}{S\xi_l}\biggr)\;,
\end{eqnarray}
where $F(x)$ is Dawson's integral function, and so the need to evaluate integrals 
in the principal-value sense is removed. With this transformation,
Equation (\ref{eq:R.last}) becomes
\begin{eqnarray} \label{eq:R.final}
R(\Omega_u,\Omega_{u'};\Omega_l,\Omega_{l'},\Omega_{f};
	\hat{\omega}_{k},\hat{\omega}_{k'};\Theta)_{\rm s.l.l.}
&=& \frac{\pi}{\Delta^2\,S\xi_l\xi_f} \nonumber \\
&&\kern -2.5in \times \Biggl\{
\exp\biggl[-\frac{(\hat\omega_k-\hat\omega_{k'}+\omega_{lf})^2}
	{\Delta^2}\biggr]\left[
W\biggl(\frac{\kappa^+ v_{ul}+\kappa^- w_{uf}}{S\xi_l\xi_f},
	\frac{a_u}{S\xi_l\xi_f} \biggr) +
\overline{W}\biggl(\frac{\kappa^+ v_{u'l}+\kappa^- w_{u'f}}{S\xi_l\xi_f},
	\frac{a_{u'}}{S\xi_l\xi_f} \biggr) \right] \nonumber \\
&&\kern -2.4in \mathop{+}
\exp\biggl[-\frac{(\hat\omega_k-\hat\omega_{k'}+\omega_{l'f})^2}
	{\Delta^2}\biggr]\left[
W\biggl(\frac{\kappa^+ v_{ul'}+\kappa^- w_{uf}}{S\xi_l\xi_f},
	\frac{a_u}{S\xi_l\xi_f} \biggr) +
\overline{W}\biggl(\frac{\kappa^+ v_{u'l'}+\kappa^- w_{u'f}}{S\xi_l\xi_f},
	\frac{a_{u'}}{S\xi_l\xi_f} \biggr) \right]
	\Biggr\} \nonumber \\
&&\kern -2.5in \,\mathop{+}
\frac{2}{\sqrt\pi}\,\frac{{\rm i}}{\Delta^2 S\xi_l\xi_f}
\int_{-\infty}^{+\infty} dp\;
{\rm e}^{-p^2/\xi_f^2} \biggl[
\frac{1}{a_u+{\rm i}(p-w_{uf})}+\frac{1}{a_{u'}-{\rm i}(p-w_{u'f})}
	\biggr] \nonumber \\
&&\kern -1.2in\times
\biggl[
F\biggl(\frac{v_{ul'}-w_{uf}+p(1-C\xi_l/\xi_f)}{S\xi_l}\biggr) -
F\biggl(\frac{v_{ul}-w_{uf}+p(1-C\xi_l/\xi_f)}{S\xi_l}\biggr)
\biggr]\;,
\end{eqnarray}
where we used Equations (\ref{eq:variables.1}) and (\ref{eq:useful}) 
in order to combine similar terms.
%

\begin{figure}[t!]
\centering
\includegraphics[height=3.7truein]{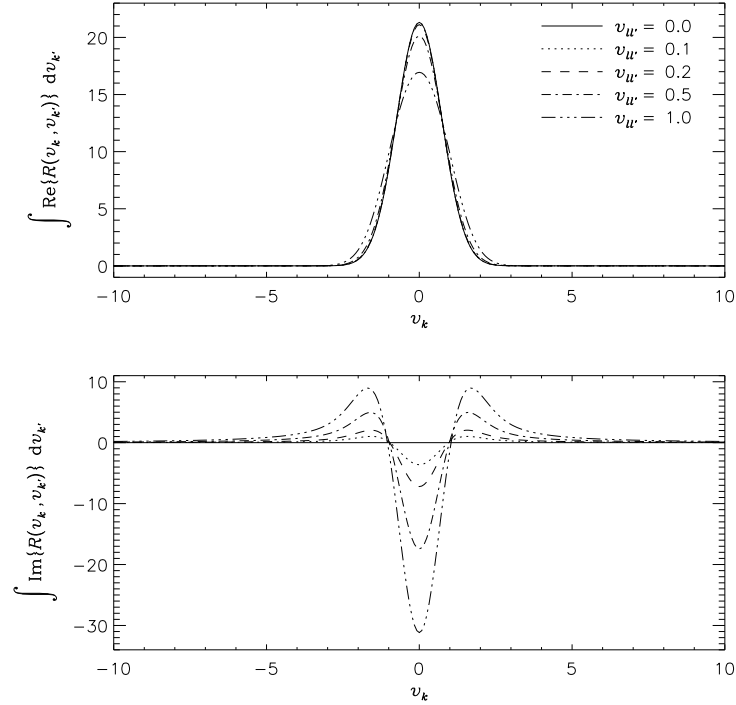}
\caption{\label{fig:int_term}
The real (top) and imaginary (bottom) parts of the integral of the 
redistribution function of Equation (\ref{eq:R.final}) over
the normalized output frequency $v_{k'}\equiv\frac{1}{2}(w_{uf}+w_{u'f})$, 
plotted against the normalized input frequency 
$v_k\equiv\frac{1}{2}(v_{ul}+v_{u'l'})$. The different 
curves correspond to different values of the ratio 
$v_{ll'}\equiv\omega_{ll'}/\Delta$. For this example, we assumed 
a scattering angle $\Theta=$1\,rad, and a fully 
degenerate upper level (i.e., $\omega_{uu'}=0$).}
\end{figure}

Owing to the fact that $|F'(x)|\le 1$ over the real domain 
\cite[e.g.,][]{AS64}, from the mean-value theorem it follows that
\begin{displaymath}
\left|F\biggl(\frac{v_{ul'}-w_{uf}+p(1-C\xi_l/\xi_f)}{S\xi_l}\biggr)
-
F\biggl(\frac{v_{ul}-w_{uf}+p(1-C\xi_l/\xi_f)}{S\xi_l}\biggr)\right|
\le
\frac{|\omega_{ll'}|}{\Delta S\xi_l}\;.
\end{displaymath}
This allows us to estimate a bound to the integral contribution of 
Equation (\ref{eq:R.final}). In fact,
\begin{eqnarray}
&&\frac{1}{\pi}\int_{-\infty}^{+\infty} dp\;
{\rm e}^{-p^2/\xi_f^2} \biggl|
\frac{1}{a_u+{\rm i}(p-w_{uf})}+\frac{1}{a_{u'}-{\rm i}(p-w_{u'f})}
	\biggr| \nonumber \\
&\le& H\!\left(\frac{w_{uf}}{\xi_f},\frac{a_u}{\xi_f}\right) +
      H\!\left(\frac{w_{u'f}}{\xi_f},\frac{a_{u'}}{\xi_f}\right) +
\frac{1}{\pi}\int_{-\infty}^{+\infty} dp\;
{\rm e}^{-p^2/\xi_f^2} \biggl[
\frac{|p-w_{uf}|}{a_u^2+(p-w_{uf})^2}
+\frac{|p-w_{u'f}|}{a_{u'}^2+(p-w_{u'f})^2}
	\biggr] \nonumber \\
\noalign{\allowbreak}
&\le& H\!\left(\frac{w_{uf}}{\xi_f},\frac{a_u}{\xi_f}\right) +
      H\!\left(\frac{w_{u'f}}{\xi_f},\frac{a_{u'}}{\xi_f}\right) +
\frac{1}{2\pi}\int_{-\infty}^{+\infty} dp\;
{\rm e}^{-p^2/\xi_f^2} \biggl( \frac{1}{a_u} + \frac{1}{a_{u'}} \biggr)
\nonumber \\
&=& H\!\left(\frac{w_{uf}}{\xi_f},\frac{a_u}{\xi_f}\right) +
      H\!\left(\frac{w_{u'f}}{\xi_f},\frac{a_{u'}}{\xi_f}\right) +
\frac{\xi_f}{2\sqrt\pi}\biggl( \frac{1}{a_u} + \frac{1}{a_{u'}} \biggr)\;,
\end{eqnarray}
and so the integral contribution in Equation (\ref{eq:R.final}) can be 
neglected when the quantity
\begin{displaymath}
\frac{2}{\sqrt\pi}\,\frac{|\omega_{ll'}|}{\Delta S\xi_l}\,
\biggl[H\!\left(\frac{w_{uf}}{\xi_f},\frac{a_u}{\xi_f}\right) +
      H\!\left(\frac{w_{u'f}}{\xi_f},\frac{a_{u'}}{\xi_f}\right) +
\frac{\xi_f}{2\sqrt\pi}\biggl( \frac{1}{a_u} + \frac{1}{a_{u'}}
\biggr)\biggr]
\end{displaymath}
is sufficiently small compared to the absolute value of the contribution 
within curly brackets in that same equation.

\begin{figure}[t!]
\centering
\includegraphics[height=3.7truein]{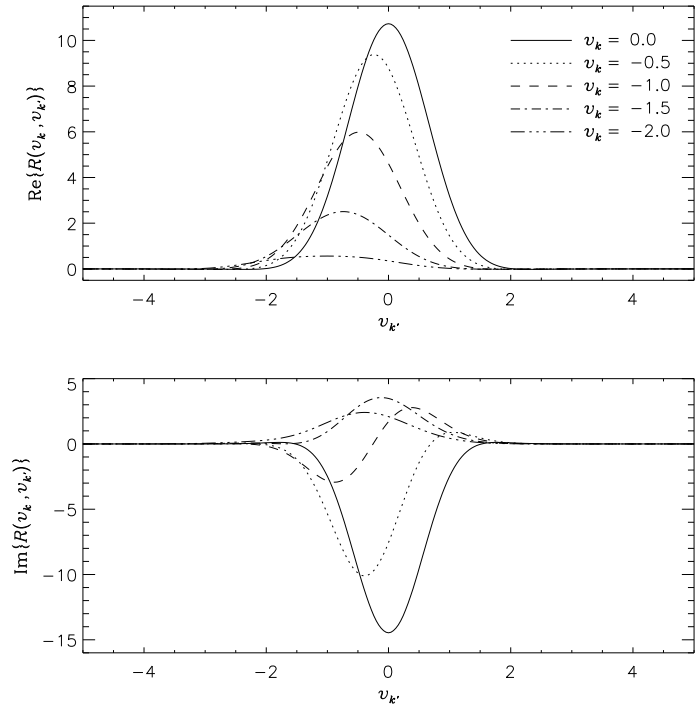}
\caption{\label{fig:Wfunc}
The real (top) and imaginary (bottom) parts of the redistribution 
function of Equation (\ref{eq:R.final}) plotted against the 
normalized output frequency $v_{k'}\equiv\frac{1}{2}(w_{uf}+w_{u'f})$, for various values of the 
normalized input frequency $v_k\equiv\frac{1}{2}(v_{ul}+v_{u'l'})$. 
For this example, we assumed $\omega_{ll'}=0.5\,\Delta$ and 
$\omega_{uu'}=0.2\,\Delta$.}
\end{figure}

Figure~\ref{fig:int_term} shows the integral over the normalized
output frequency $v_{k'}\equiv\frac{1}{2}(w_{uf}+w_{u'f})$ of the 
redistribution function of Equation (\ref{eq:R.final}), plotted 
against the normalized input frequency
$v_k\equiv\frac{1}{2}(v_{ul}+v_{u'l'})$. The upper (lower) 
panel shows the real (imaginary) part of this integral. The different 
curves correspond to different values of the quantity 
$v_{ll'}\equiv\omega_{ll'}/\Delta$. For this example, we assumed a
scattering angle $\Theta=1$\,rad, and a fully degenerate upper 
state, so that $\omega_{uu'}=0$. In
such a case, we see from Equation (\ref{eq:R.final}) that the
contribution within curly brackets is purely real. Therefore, the plots
in the lower panel are exclusively due to the integral term in
Equation (\ref{eq:R.final}). We can conclude from those plots
that the contribution of 
the integral term is generally not negligible, so it can significantly 
affect the shape of the redistributed profile depending on the
importance of the lower-term coherence.
In contrast, the integral term brings no contribution when the atom is 
illuminated with a flat spectrum, owing to the fact that its integral 
over $v_k$ vanishes identically (see Figure~\ref{fig:int_term}, and 
the comment at the end of Appendix~\ref{app:C}).

In the particular case when the atomic coherence of the lower term 
is completely relaxed ($\rho_{ll'}=\delta_{ll'}\,\rho_{ll}$), or if the
levels $l$ and $l'$ are completely degenerate, the integral contribution
vanishes, and Equation (\ref{eq:R.final}) provides the 
generalization of the well-known $R_{\rm II}$ redistribution function 
\citep[e.g.,][]{Mi78} to the case of a $\Lambda$-type three-term atom,
\begin{eqnarray}	\label{eq:R2}
R_{\rm II}(\Omega_u,\Omega_{u'};\Omega_l,\Omega_{f};
	\hat{\omega}_{k},\hat{\omega}_{k'};\Theta)
&=& \frac{2\pi}{\Delta^2\,S\xi_l\xi_f}\,
\exp\biggl[-\frac{(\hat\omega_k-\hat\omega_{k'}+\omega_{lf})^2}
	{\Delta^2}\biggr] \nonumber \\
&&\kern -3cm {}\times \biggl[
W\biggl(\frac{\kappa^+ v_{ul}+\kappa^- w_{uf}}{S\xi_l\xi_f},
	\frac{a_u}{S\xi_l\xi_f} \biggr) +
\overline{W}\biggl(\frac{\kappa^+ v_{u'l}+\kappa^- w_{u'f}}{S\xi_l\xi_f},
	\frac{a_{u'}}{S\xi_l\xi_f} \biggr) \biggr]\;.
\end{eqnarray}
This expression was applied in recent work that investigated the effects
of partial redistribution on the formation of polarized lines from
$\Lambda$-type three-term atoms in the solar spectrum \citep{CM16}.

Finally, Figure~\ref{fig:Wfunc} shows several realizations of the full
redistribution function of Equation (\ref{eq:R.final}) plotted against
the normalized output frequency $v_{k'}$, for different values 
of the normalized input frequency $v_k$.
For the example in this figure, we assumed 
$\omega_{ll'}=0.5\,\Delta$ and 
$\omega_{uu'}=0.2\,\Delta$, and a scattering angle $\Theta=1$\,rad.

\section{The cases of forward and backward scattering}
\label{sec:limitcase}

In the two cases of forward and backward scattering 
($\Theta=0$ and $\Theta=\pi$, respectively), the expression
(\ref{eq:R.final}) for the laboratory redistribution function breaks
down because of the condition $S=0$. We can, however, treat these two
cases relying on a physical argument of continuity, and taking the 
limit of the redistribution function for $S\to 0^+$.
Using the asymptotic expansion of $W(v,a)$ for large values of 
$|v+{\rm i}a|$ \citep[e.g.,][]{LL04}, and recalling 
Equation (\ref{eq:F_def}), it can be shown that
\begin{equation} \label{eq:asympt}
\lim_{k\to 0^+} \frac{1}{k}\,W\Bigl(\frac{v}{k},\frac{a}{k}\Bigr)
	=\frac{{\rm i}}{\sqrt{\pi}\,(v+{\rm i}a)}\;,	\qquad
\lim_{k\to 0^+} \frac{1}{k}\,F\Bigl(\frac{v}{k}\Bigr)
	=\frac{1}{2v}\;.
\end{equation}
Considering for simplicity the case of non-coherent lower term, 
Equation (\ref{eq:R2}) thus becomes, for the two cases of forward and
backward scattering,
\begin{eqnarray} \label{eq:R2.fb}
R_{\rm II}(\Omega_u,\Omega_{u'};\Omega_l,\Omega_{f};
	\hat{\omega}_{k},\hat{\omega}_{k'};C=\pm1)
&=& \frac{2\sqrt\pi}{\Delta_\mp}\,
\exp\biggl[-\frac{(\hat\omega_k-\hat\omega_{k'}+\omega_{lf})^2}
	{\Delta_\mp^2}\biggr] \nonumber \\
&&\kern -2.5in {}\times \biggl[
\frac{{\rm i}}{\kappa^+ (\hat\omega_k-\omega_{ul})
+\kappa^- (\hat\omega_{k'}-\omega_{uf})+{\rm i}\epsilon_u} -
\frac{{\rm i}}{\kappa^+ (\hat\omega_k-\omega_{u'l})
+\kappa^- (\hat\omega_{k'}-\omega_{u'f})-{\rm i}\epsilon_{u'}}
\biggr]\;,
\end{eqnarray}
where $\Delta_\pm=\Delta_{ul}\pm\Delta_{uf}$.


\begin{figure}[t!]
\centering
\includegraphics[height=3.7truein]{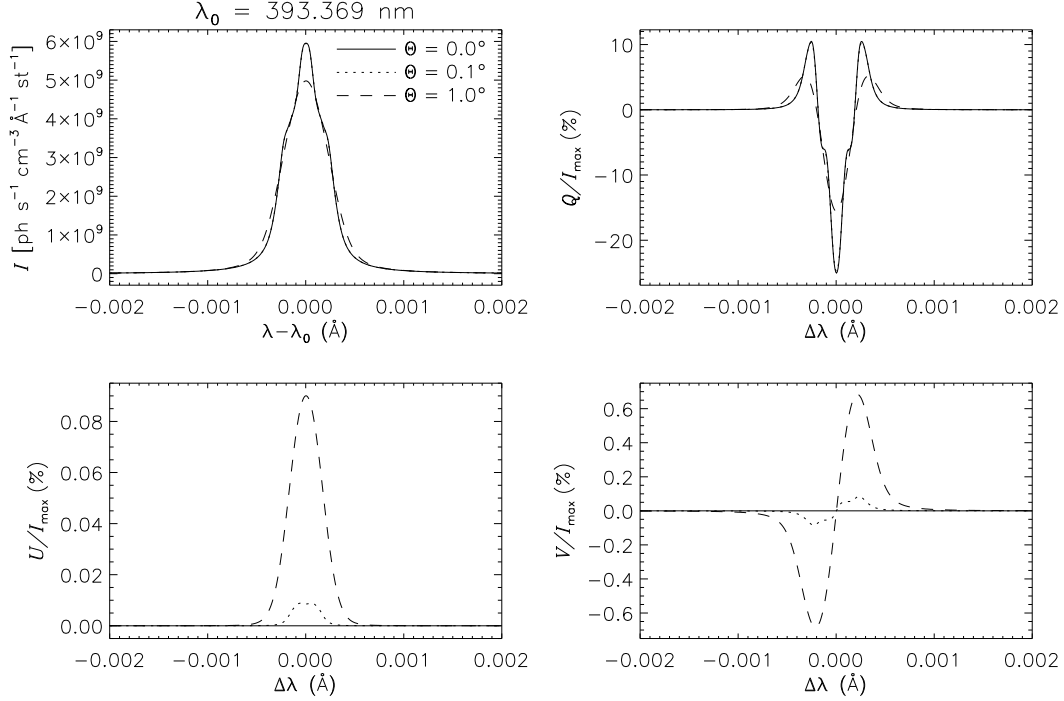}
\caption{\label{fig:FS test}
Plots showing the transition between the forms (\ref{eq:R2}) and
(\ref{eq:R2.fb}) of $R_{\rm II}$ in approaching the condition of forward 
scattering. The atomic model adopted is that of the three-term \ion{Ca}{2} 
ion encompassing the formation of the H and K lines and the IR triplet. 
The plots show the Raman scattered emissivity in the polarized 
\ion{Ca}{2} K line due to the monochromatic excitation of the 
\ion{Ca}{2} 854.2\,nm line at the exact value of its resonance 
frequency, and for different values of the scattering angle $\Theta$ 
near $0^\circ$.}
\end{figure}

\subsection{The Special Case of the Two-term Atom}

When the two lower terms coincide, so that $\Delta_{uf}=\Delta_{ul}$,
the asymptotic result expressed 
by Equation (\ref{eq:asympt}) is not generally applicable. In fact, 
in the case of a two-term atom
with non-coherent lower term, Equation (\ref{eq:R2}) becomes
\begin{eqnarray} \label{eq:R2.two}
R_{\rm II}(\Omega_u,\Omega_{u'};\Omega_l,\Omega_{f};
	\hat{\omega}_{k},\hat{\omega}_{k'};\Theta)
&=&\frac{\pi}{\Delta\omega_T^2\,C_2 S_2} 
\exp\!\left[-\frac{(\hat{\omega}_k-\hat{\omega}_{k'}+\omega_{lf})^2}%
	{4 S_2^2\Delta\omega_T^2}\right] \nonumber \\
&&\kern -2in {}\times
\left[
W\biggl(\frac{\hat{\omega}_k+\hat{\omega}_{k'}
	-\omega_{ul}-\omega_{uf}}{2 C_2\Delta\omega_T},
	\frac{\epsilon_u}{C_2\Delta\omega_T}\biggr) +
\overline{W}\biggl(\frac{\hat{\omega}_k+\hat{\omega}_{k'}
	-\omega_{u'l}-\omega_{u'f}}{2 C_2\Delta\omega_T},
	\frac{\epsilon_{u'}}{C_2\Delta\omega_T}\biggr)\right]\;,
\end{eqnarray}
where we indicated with $\Delta\omega_T=\Delta_{ul}=\Delta_{uf}$ 
the single value of the Doppler width of the two-term transition, 
and we also defined
$S_2=\sin(\Theta/2)$ and $C_2=\cos(\Theta/2)$ (cf.\ Paper~I).
Then the limit (\ref{eq:asympt}) applies only to the case of backward 
scattering $(C_2\to 0)$, but not to forward scattering $(S_2\to 0)$.
In the first case, we obtain an expression formally identical to
Equation (\ref{eq:R2.fb}), namely,
\begin{eqnarray} \label{eq:R2.two.backward}
R_{\rm II}(\Omega_u,\Omega_{u'};\Omega_l,\Omega_{f};
	\hat{\omega}_{k},\hat{\omega}_{k'};\Theta=\pi)
&=& \frac{2\sqrt\pi}{\Delta\omega_T}\,
\exp\biggl[-\frac{(\hat\omega_k-\hat\omega_{k'}+\omega_{lf})^2}
	{4\Delta\omega_T^2}\biggr] \nonumber \\
&&\kern -2in {}\times \Biggl(
\frac{{\rm i}}{\hat{\omega}_k+\hat{\omega}_{k'}
	-\omega_{ul}-\omega_{uf}+2{\rm i}\epsilon_u} -
\frac{{\rm i}}{\hat{\omega}_k+\hat{\omega}_{k'}
	-\omega_{u'l}-\omega_{u'f}-2{\rm i}\epsilon_{u'}}
\Biggr)\;,
\end{eqnarray}
whereas in the case of forward scattering, 
\begin{eqnarray} \label{eq:R2.two.forward}
R_{\rm II}(\Omega_u,\Omega_{u'};\Omega_l,\Omega_{f};
	\hat{\omega}_{k},\hat{\omega}_{k'};\Theta=0)
&=&\frac{2\pi\sqrt\pi}{\Delta\omega_T}\,
	\delta(\hat{\omega}_k-\hat{\omega}_{k'}+\omega_{lf})
	\nonumber \\
&&\kern -1.5in {}\times
\left[
W\biggl(\frac{\hat{\omega}_{k'}-\omega_{uf}}{\Delta\omega_T},
	\frac{\epsilon_u}{\Delta\omega_T}\biggr) +
\overline{W}\biggl(\frac{\hat{\omega}_{k'}-\omega_{u'f}}{\Delta\omega_T},
	\frac{\epsilon_{u'}}{\Delta\omega_T}\biggr)\right]\;,
\end{eqnarray}
owing to the fact that $\exp(-x^2/k^2)/k\to\sqrt{\pi}\,\delta(x)$ when 
$k\to0^+$.
This last expression in particular shows that the process of forward 
scattering \emph{in a two-term atom} is strictly coherent.
In fact, from a simple inspection of Equation (\ref{eq:R2.fb}) it is 
already possible to 
conclude that both processes of forward and backward scattering imply an 
increased degree of correlation between the input and output 
frequencies \citep[cf.][and references therein]{Le77},
with a typical spread that is dominated by the inverse
lifetime of the upper level, rather than by the Doppler width. However,
only in the case of forward scattering in a two-term atom, strict coherence
is attained, according to Equation (\ref{eq:R2.two.forward}).

The plots in Figure~\ref{fig:FS test} show an example of the behavior of
the scattered polarized emissivity in a three-term atomic system when the
scattering angle approaches $0^\circ$. For the example in
this figure, we adopted the three-term model of the \ion{Ca}{2} ion
underlying the formation of the H and K doublet around 395\,nm and the 
IR triplet around 858\,nm. The line formation model also includes a
magnetic field of 10\,G normal to the direction of the incident light, 
which is responsible for the appearance of line polarization, and in 
particular of non-vanishing Stokes $U$ and $V$ signals when the 
scattering direction forms an angle with the incidence direction. 
The plotted profiles of the 
\ion{Ca}{2} K line at 393.4\,nm are produced by Raman scattering of 
monochromatic radiation at the exact resonance frequency of the 854.2\,nm 
line of the IR triplet. For the modeling we assumed a plasma density of 
$10^{12}\,\rm cm^{-3}$ and a temperature of 1000\,K, which corresponds to 
a Doppler width of 8.5\,m\AA. However, because of the near condition of 
forward scattering, the actual width of the scattered profiles is  
instead dominated by the natural linewidth ($\sim 0.5$\,m\AA),
as expected.
Figure~\ref{fig:FS test} numerically demonstrates how 
Equation (\ref{eq:R2.fb}) indeed provides the correct limit of the 
general expression (\ref{eq:R2}) of $R_{\rm II}$, when approaching 
the condition of forward scattering, as both Stokes $I$ and $Q$ for 
$\Theta=0.1^\circ$ are already practically indistinguishable from 
those at $\Theta=0^\circ$.


\begin{acknowledgements}
We thank T.\ del Pino Alem\'an (HAO) for internally reviewing the 
manuscript and for helpful comments. We are deeply indebted 
to the anonymous referee, for a very careful review of our manuscript and
for helpful comments and suggestions that have greatly improved the 
presentation of this work.
\end{acknowledgements}

\appendix

\section{Formal limit for the condition of metastable initial and final 
terms} \label{app:A}

In order to derive the form of Equation (\ref{eq:rlab.bar}) for the case 
of metastable initial and final states, we must consider two possibilities 
in taking the limit for $\bar a_{l,l',f}\to 0$, which correspond to the 
assumptions that $\bar a_f \gtrless \bar a_{l,l'}$. We then find, in
that limit,
\begin{eqnarray}
R(\Omega_u,\Omega_{u'};\Omega_l,\Omega_{l'},\Omega_{f};
	\hat{\omega}_{k},\hat{\omega}_{k'};\Theta)_{\rm s.l.l.} &=&
\frac{1}{\Delta^2\,S\xi_l\xi_f} 
\int_{-\infty}^{+\infty} d\bar q\; {\rm e}^{-S^2\xi_l^2\xi_f^2\bar q^2}
 \nonumber \\
&&\kern -2.5in\times\Biggl\{
W\bigl( x_{ul'},\bar a_u\bigr)
\biggl[
\pi\delta(x_{ul'}-y_{uf})
+{\rm i}\,{\rm Pv}\frac{1}{x_{ul'}-y_{uf}}\biggr]
+\overline{W}\bigl( x_{u'l},\bar a_{u'}\bigr)
\biggl[
\pi\delta(x_{u'l}-y_{u'f})
-{\rm i}\,{\rm Pv}\frac{1}{x_{u'l}-y_{u'f}}\biggr] \nonumber \\
&&\kern -2.5in \,\,{}+
W\bigl( y_{uf},\bar a_u\bigr)
\biggl[
\pi\delta(x_{ul}-y_{uf})
-{\rm i}\,{\rm Pv}\frac{1}{x_{ul}-y_{uf}}\biggr]
+\overline{W}\bigl( y_{u'f},\bar a_{u'}\bigr)
\biggl[
\pi\delta(x_{u'l'}-y_{u'f})
+{\rm i}\,{\rm Pv}\frac{1}{x_{u'l'}-y_{u'f}}\biggr] \nonumber \\
\noalign{\allowbreak}
&&\kern -2.5in \,\,{}+
\Bigl[W\bigl( x_{ul'},\bar a_u\bigr)
	- W\bigl( y_{uf},\bar a_u\bigr)\Bigr]
\biggl[
\pm\pi\delta(x_{ul'}-y_{uf})
-{\rm i}\,{\rm Pv}\frac{1}{x_{ul'}-y_{uf}}\biggr] \nonumber \\
&&\kern -2.5in \,\,{}+
\Bigl[\overline{W}\bigl( x_{u'l},\bar a_{u'}\bigr)
	-\overline{W}\bigl( y_{u'f},\bar a_{u'}\bigr)\Bigr]
\biggl[
\pm\pi\delta(x_{u'l}-y_{u'f})
+{\rm i}\,{\rm Pv}\frac{1}{x_{u'l}-y_{u'f}}\biggr]
\Biggr\}\;,
\end{eqnarray}
where the $\pm$ sign in front of the Dirac-$\delta$s in the last two
lines must be chosen according to whether $\bar a_f\gtrless \bar a_l$.
By noting that
\begin{eqnarray*}
W\bigl( y_{uf},\bar a_u\bigr)\,\delta(x_{ul'}-y_{uf})
&=& W\bigl( x_{ul'},\bar a_u\bigr)\,\delta(x_{ul'}-y_{uf})\;,
\\
\overline{W}\bigl( y_{u'f},\bar a_{u'}\bigr)\,\delta(x_{u'l}-y_{u'f})
&=& \overline{W}\bigl( x_{u'l},\bar
a_{u'}\bigr)\,\delta(x_{u'l}-y_{u'f})\;,
\end{eqnarray*}
we see that the terms carrying a sign choice vanish identically, and so
we are left with
\begin{eqnarray} \label{eq:R.temp}
R(\Omega_u,\Omega_{u'};\Omega_l,\Omega_{l'},\Omega_{f};
	\hat{\omega}_{k},\hat{\omega}_{k'};\Theta)_{\rm s.l.l.} &=&
\frac{\pi}{\Delta^2\,S\xi_l\xi_f} 
\int_{-\infty}^{+\infty} d\bar q\; {\rm e}^{-S^2\xi_l^2\xi_f^2\bar q^2} 
	\nonumber \\
&&\kern -2.2in\times\Biggl\{
W\bigl( x_{ul'},\bar a_u\bigr)\,\delta(x_{ul'}-y_{uf})
+
\overline{W}\bigl( x_{u'l},\bar a_{u'}\bigr)\,\delta(x_{u'l}-y_{u'f})
+W\bigl( y_{uf},\bar a_u\bigr)\,\delta(x_{ul}-y_{uf})
+
\overline{W}\bigl( y_{u'f},\bar a_{u'}\bigr)\,\delta(x_{u'l'}-y_{u'f})
\nonumber \\
&&\kern -2.2in \,\,{}
+ \frac{\rm i}{\pi}\,
W\bigl( y_{uf},\bar a_u\bigr)
\biggl({\rm Pv}\frac{1}{x_{ul'}-y_{uf}}-{\rm
Pv}\frac{1}{x_{ul}-y_{uf}}\biggr) 
+ \frac{\rm i}{\pi}\,
\overline{W}\bigl( y_{u'f},\bar a_{u'}\bigr)
\biggl({\rm Pv}\frac{1}{x_{u'l'}-y_{u'f}}-{\rm
Pv}\frac{1}{x_{u'l}-y_{u'f}}\biggr)
\Biggr\}\;,
\end{eqnarray}
We now substitute back the relations (\ref{eq:x.y.def}) and
(\ref{eq:ganzo}) that hold for the variables $x_{ab}$ and $y_{ab}$:
\begin{eqnarray} \label{eq:R.temp1}
R(\Omega_u,\Omega_{u'};\Omega_l,\Omega_{l'},\Omega_{f};
	\hat{\omega}_{k},\hat{\omega}_{k'};\Theta)_{\rm s.l.l.} 
&=&
\frac{\pi}{\Delta^2\,S\xi_l\xi_f} 
\int_{-\infty}^{+\infty} d\bar q\; {\rm e}^{-S^2\xi_l^2\xi_f^2\bar q^2} 
	\nonumber \\
&&\kern -2.2in\times\Bigr[\,
W\bigl( \bar v_{ul'}+\bar q(C\xi_l\xi_f-\xi_l^2),
	\bar a_u \bigr)\,
	\delta(\bar v_{ul'}-\bar w_{uf}-\bar q\bigr)
+\overline{W}\bigl( \bar v_{u'l}+\bar q(C\xi_l\xi_f-\xi_l^2),
	\bar a_{u'} \bigr)\,
	\delta(\bar v_{u'l}-\bar w_{u'f}-\bar q\bigr) 
	\nonumber \\
&&\kern -2.2in \,{}+
W\bigl( \bar w_{uf}+\bar q(\xi_f^2-C\xi_l\xi_f),
	\bar a_u \bigr)\,
	\delta(\bar v_{ul}-\bar w_{uf}-\bar q \bigr)
+\overline{W}\bigl( \bar w_{u'f}+\bar q(\xi_f^2-C\xi_l\xi_f),
	\bar a_{u'} \bigr)\,
	\delta(\bar v_{u'l'}-\bar w_{u'f}-\bar q \bigr)\Bigr]
	\nonumber \\
\noalign{\allowbreak}
&&\kern -2.35in{}+ \frac{\rm i}{\Delta^2\,S\xi_l\xi_f} 
\int_{-\infty}^{+\infty} d\bar q\; {\rm e}^{-S^2\xi_l^2\xi_f^2\bar q^2} 
\Biggl\{
W\bigl( \bar w_{uf}+\bar q(\xi_f^2-C\xi_l\xi_f),
	\bar a_u \bigr) \biggl(
{\rm Pv}\frac{1}{\bar v_{ul'}-\bar w_{uf}-\bar q}-
{\rm Pv}\frac{1}{\bar v_{ul}-\bar w_{uf}-\bar q}
	\biggr) \nonumber \\
&&\kern -.35in{}+
\overline{W}\bigl( \bar w_{u'f}+\bar q(\xi_f^2-C\xi_l\xi_f),
	\bar a_{u'} \bigr) \biggl(
{\rm Pv}\frac{1}{\bar v_{u'l'}-\bar w_{u'f}-\bar q}-
{\rm Pv}\frac{1}{\bar v_{u'l}-\bar w_{u'f}-\bar q}
	\biggr) \Biggr\}
\nonumber \\
\noalign{\allowbreak}
&&\kern -2.35in \,\mathop{\equiv}
\frac{\pi}{\Delta^2\,S\xi_l\xi_f} \Biggl\{
W\biggl(\frac{\kappa^+ v_{ul'}+\kappa^- w_{uf}}{S\xi_l\xi_f},
	\frac{a_u}{S\xi_l\xi_f} \biggr)\,
\exp\!\left[- (v_{ul'}-w_{uf})^2\right] 
+\overline{W}\biggl(\frac{\kappa^+ v_{u'l}+\kappa^- w_{u'f}}{S\xi_l\xi_f},
	\frac{a_{u'}}{S\xi_l\xi_f} \biggr)\,
\exp\!\left[- (v_{u'l}-w_{u'f})^2\right]
	\nonumber \\
&&\kern -1.85in {}+
W\biggl(\frac{\kappa^+ v_{ul}+\kappa^- w_{uf}}{S\xi_l\xi_f},
	\frac{a_u}{S\xi_l\xi_f} \biggr)\,
\exp\!\left[- (v_{ul}-w_{uf})^2\right]
+\overline{W}\biggl(\frac{\kappa^+ v_{u'l'}+\kappa^- w_{u'f}}{S\xi_l\xi_f},
	\frac{a_{u'}}{S\xi_l\xi_f} \biggr)\,
\exp\!\left[- (v_{u'l'}-w_{u'f})^2\right]
	\Biggr\} \nonumber \\
\noalign{\allowbreak}
&&\kern -2.35in \,\mathop{+}
\frac{\rm i}{\Delta^2\,S\xi_l\xi_f} 
\fint_{-\infty}^{+\infty} dq\; {\rm e}^{-q^2} 
\Biggl\{
W\biggl( \frac{w_{uf} + q(\xi_f^2-C\xi_l\xi_f)}{S\xi_l\xi_f},
	\frac{a_u}{S\xi_l\xi_f} \biggr)
\biggl(
\frac{1}{v_{ul'}-w_{uf}-q}-
\frac{1}{v_{ul}-w_{uf}-q}
	\biggr) \nonumber \\
&&\kern -1.in \,\,{}+
\overline{W}\biggl( \frac{w_{u'f} + q(\xi_f^2-C\xi_l\xi_f)}{S\xi_l\xi_f},
	\frac{a_{u'}}{S\xi_l\xi_f} \biggr)
\biggl(
\frac{1}{v_{u'l'}-w_{u'f}-q}-
\frac{1}{v_{u'l}-w_{u'f}-q}
	\biggr) \Biggr\}\;,
\end{eqnarray}
where we introduced new quantities (cf.\ Equation (\ref{eq:cool}))
\begin{equation} \label{eq:chi}
\kappa^\pm={\textstyle\frac{1}{2}}\bigl[1\pm(\xi_f^2-\xi_l^2)\bigr]\;,
\end{equation}
and where the symbol $\tfint$ in the last part indicates that the 
integral is taken in the principal value sense.
%

Finally, by recalling the definition (\ref{eq:variables.1}) for the 
quantities $v_{ab}$ and $w_{ab}$, after reordering of some 
terms, we arrive at Equation (\ref{eq:R.last}).

\section{Transformation of the integral term}
\label{app:B}

We consider the expression
\begin{equation} \label{eq:term}
{\cal T}=\fint_{-\infty}^{+\infty} dq\;
{\rm e}^{-q^2}\,\frac{W((w+\beta q)/N,a/N)}
{v-w-q}\;,
\end{equation}
with $N\equiv S\xi_l\xi_f$ and $\beta\equiv\xi_f^2 -C\xi_l\xi_f$. This
expression provides the generic form of the various contributions to the
integral term of Equation (\ref{eq:R.last}).
Using the explicit form of $W(v,a)$, Equation (\ref{eq:W}), we find
\begin{eqnarray*}
{\cal T}&=&
\frac{1}{\pi}
\fint_{-\infty}^{+\infty} dq
\int_{-\infty}^{+\infty} dp\;
\frac{{\rm e}^{-p^2}}
{a/N+{\rm i}[p-(w+\beta q)/N]}\,
\frac{{\rm e}^{-q^2}}{v-w-q} \\
&=&\frac{1}{\pi}
\fint_{-\infty}^{+\infty} dq
\int_{-\infty}^{+\infty} dp\;
\frac{{\rm e}^{-(p+\beta q/N)^2}}
{a/N+{\rm i}(p-w/N)}\,
\frac{{\rm e}^{-q^2}}{v-w-q}\;,
\end{eqnarray*}
where in the second line we operated the substitution $p\to p+\beta q/N$.
We note that
\begin{eqnarray*}
\biggl(p+\frac{\beta q}{N}\biggr)^2+q^2
&=&p^2+2\frac{\beta}{N}\,p q
+\biggl(1+\frac{\beta^2}{N^2}\biggr)q^2 
\equiv p^2+2\frac{\beta}{N}\,p q
+\frac{q^2}{\alpha^2} \\
&=&\biggl(1-\frac{\alpha^2\beta^2}{N^2}\biggr)p^2+
\biggl(\frac{\alpha\beta}{N}\,p+\frac{q}{\alpha}\biggr)^2 
=\alpha^2 p^2+
\biggl(\frac{\alpha\beta}{N}\,p+\frac{q}{\alpha}\biggr)^2\;,
\end{eqnarray*}
where we put $\alpha^2\equiv (1+\beta^2/N^2)^{-1}=S^2\xi_l^2$
(see Equation (\ref{eq:cool})). 
Defining next $\tilde q=q/\alpha$, we find at last
\begin{eqnarray} \label{eq:transform}
{\cal T}
&=&\frac{1}{\pi}
\int_{-\infty}^{+\infty} dp\;
\frac{{\rm e}^{-\alpha^2 p^2}}
{a/N+{\rm i}(p-w/N)}
\fint_{-\infty}^{+\infty} d\tilde q\;
\frac{{\rm e}^{-(\tilde q+\alpha\beta p/N)^2}}
{(v-w)/\alpha+\alpha\beta p/N -(\tilde q+\alpha\beta p/N)} 
	\nonumber \\
&\equiv&
\frac{2}{\sqrt{\pi}}
\int_{-\infty}^{+\infty} dp\;
\frac{{\rm e}^{-\alpha^2 p^2}}
{a/N+{\rm i}(p-w/N)}\,
F\biggl(\frac{v-w}{\alpha}+\frac{\alpha\beta}{N}\,p\biggr)
	\nonumber \\
&=&
\frac{2}{\sqrt{\pi}}
\int_{-\infty}^{+\infty} dp\;
\frac{{\rm e}^{-(\alpha/N)^2 p^2}}
{a+{\rm i}(p-w)}\,
F\biggl(\frac{v-w}{\alpha}+\frac{\alpha\beta}{N^2}\,p\biggr)
	\nonumber \\
&=&
\frac{2}{\sqrt{\pi}}
\int_{-\infty}^{+\infty} dp\;
\frac{{\rm e}^{-p^2/\xi_f^2}}{a+{\rm i}(p-w)}\,
F\biggl(\frac{v-w+p(1-C\xi_l/\xi_f)}{S\xi_l}\biggr)\;,
\end{eqnarray}
where in the second line we recognized the integral over $\tilde q$
to be the Hilbert transform of a Gaussian, which can therefore be 
expressed in terms of the Dawson's integral function $F(x)$ 
\cite[e.g.,][]{GW01}.

Finally, we observe that
\begin{equation} \label{eq:F_def}
F(v)=\frac{\sqrt\pi}{2}\,L(v,0)\;,
\end{equation}
where $L(v,a)$ is the 
imaginary part of the generalized Voigt function $W(v,a)$ of 
Equation (\ref{eq:W}). Hence the expression (\ref{eq:transform})
can conveniently be evaluated using the same algorithm adopted for 
the computation of $W(v,a)$.

\section{Normalization of the redistribution function}
\label{app:C}

It is important to verify that the transformation of the redistribution
function from the atomic frame of rest to the laboratory system preserves
its norm, which also increases confidence in the correctness of the 
transformation itself.
In order to do so, we consider the simplest case of 
Equation (\ref{eq:R2}), which we rewrite in the form
(see Equation (\ref{eq:useful}))
\begin{eqnarray}
R_{\rm II}(\Omega_u,\Omega_{u'};\Omega_l,\Omega_{f};
	\hat{\omega}_{k},\hat{\omega}_{k'};\Theta)
&=& \frac{2\pi}{\Delta^2\,S\xi_l\xi_f} \nonumber \\
&&\kern -2.0in {}\times \Biggl\{
\exp\!\left[-(v_{ul}-w_{uf})^2\right]
W\biggl(\frac{\kappa^+ v_{ul}+\kappa^- w_{uf}}{S\xi_l\xi_f},
	\frac{a_u}{S\xi_l\xi_f} \biggr) 
+\exp\!\left[-(v_{u'l}-w_{u'f})^2\right]
 \overline{W}\biggl(\frac{\kappa^+ v_{u'l}+\kappa^- w_{u'f}}{S\xi_l\xi_f},
	\frac{a_{u'}}{S\xi_l\xi_f} \biggr) \Biggr\}\;.
\end{eqnarray}
By introducing new variables $\xi=v_{ab}-w_{ac}$ and
$\eta=\kappa^+v_{ab}+\kappa^-w_{ac}$, we can show that the double 
integral of this expression over $\hat{\omega}_{k}$ and 
$\hat{\omega}_{k'}$ is equivalent to (we note that $\kappa^\pm$ are
dimensionless quantities, and that $\kappa^++\kappa^-=1$; 
cf.\ Equation (\ref{eq:chi}))
\begin{eqnarray*}
\int_{-\infty}^{+\infty}d\hat{\omega}_{k}
\int_{-\infty}^{+\infty}d\hat{\omega}_{k'}\;
R_{\rm II}(\Omega_u,\Omega_{u'};\Omega_l,\Omega_{f};
	\hat{\omega}_{k},\hat{\omega}_{k'};\Theta)
&=&\frac{2\pi\sqrt{\pi}}{S\xi_l\xi_f} \\
&&\kern -3.5in {}\times
\lim_{M\to\infty} \Biggl\{
\int_{-M}^0 d\eta\;\frac{1}{2}\biggl[
\mbox{erf}\left(\frac{M+2\kappa^+\eta}
	{2\kappa^+\kappa^-}\right)+
\mbox{erf}\left(\frac{M+2\kappa^-\eta}
	{2\kappa^+\kappa^-}\right)
\biggr] 
\left[W\left(\frac{\eta}{S\xi_l\xi_f},\frac{a_u}{S\xi_l\xi_f}\right) +
\overline{W}\left(\frac{\eta}{S\xi_l\xi_f},\frac{a_{u'}}{S\xi_l\xi_f}
\right)\right] \\
&&\kern -3.5in \hphantom{\lim_{M\to\infty} \Biggl\{\,}+
\int_0^M d\eta\;\frac{1}{2}\biggl[
\mbox{erf}\left(\frac{M-2\kappa^+\eta}
	{2\kappa^+\kappa^-}\right)+
\mbox{erf}\left(\frac{M-2\kappa^-\eta}
	{2\kappa^+\kappa^-}\right)
\biggr] 
\left[W\left(\frac{\eta}{S\xi_l\xi_f},\frac{a_u}{S\xi_l\xi_f}\right) +
\overline{W}\left(\frac{\eta}{S\xi_l\xi_f},\frac{a_{u'}}{S\xi_l\xi_f}
\right)\right]\Biggr\}\;.
\end{eqnarray*}
For large positive values of $k$, $\mbox{erf}(k\pm v)\sim 1$ except 
when $v$ lies within some neighborhood of $\mp k$ (in which cases 
$\mbox{erf}(k\pm v)\sim\mbox{erf}(0)=0$), whereas the function 
$W(v,a)$ tends to zero for large arguments of $|v|$. Therefore, by 
choosing a large enough value of $M$, we can replace all instances of 
$\mbox{erf}(x)$ in the above integrals with 1, since the regions where 
that function differs from 1 can be made to correspond to regions where 
the function $W(v,a)$ practically vanishes. Hence,
\begin{eqnarray*}
\int_{-\infty}^{+\infty}d\hat{\omega}_{k}
\int_{-\infty}^{+\infty}d\hat{\omega}_{k'}\;
R_{\rm II}(\Omega_u,\Omega_{u'};\Omega_l,\Omega_{f};
	\hat{\omega}_{k},\hat{\omega}_{k'};\Theta) 
&=&\frac{2\pi\sqrt{\pi}}{S\xi_l\xi_f}
\int_{-\infty}^{+\infty} d\eta
\left[W\left(\frac{\eta}{S\xi_l\xi_f},\frac{a_u}{S\xi_l\xi_f}\right) +
\overline{W}\left(\frac{\eta}{S\xi_l\xi_f},\frac{a_{u'}}{S\xi_l\xi_f}
\right)\right] \\
&=&\frac{2\pi\sqrt{\pi}}{S\xi_l\xi_f}\left(2\sqrt{\pi}\,S\xi_l\xi_f\right)
=4\pi^2\;,
\end{eqnarray*}
owing to the fact that the norm of both $W(v,a)$ and $\overline{W}(v,a)$
evaluates to $\sqrt{\pi}$.
This result is in agreement with Equation (16) of Paper~I. In 
particular, it shows that the additional integral term in 
Equation (\ref{eq:R.final}), which is generally non-vanishing in the 
presence of lower-term coherence, is purely a term of frequency 
redistribution that carries no net energy. This can also be seen 
directly from Equation (\ref{eq:R.final}), by integrating that term 
over $\hat\omega_k$. Since that variable only appears in the argument 
of the Dawson's integral, which is an odd function, the integration 
over $\hat\omega_k$ makes each of the two contributions in that term 
to vanish.
In particular this implies that the integral term of
Equation (\ref{eq:R.final}) brings no contribution when the atom is
illuminated by a spectrally flat radiation.
Figure~\ref{fig:int_term} also provides a numerical demonstration of 
this result, since the integrals over $v_k$ of the curves in the upper 
panel all converge to $4\pi^2$, whereas the curves in the lower panel 
all have zero integral.


\begin{thebibliography}

\bibitem[\protect\citeauthoryear{Abramowitz \& Stegun}{1964}]{AS64}
Abramowitz, M., \& Stegun, I.~A.~1964, Handbook of Mathematical
Functions (Washington: National Bureau of Standards)

\bibitem[\protect\citeauthoryear{Belluzzi \& Trujillo Bueno}{2012}]{BT12}
Belluzzi, L., \& Trujillo Bueno, J.~2012, \apjl, 750, L11

\bibitem[\protect\citeauthoryear{Casini \& Manso Sainz}{2016}]{CM16}
Casini, R., \& Manso Sainz, R..~2016, \apj, 824, 135

\bibitem[\protect\citeauthoryear{Casini et al.}{2014}]{Ca14}
Casini, R., Landi Degl'Innocenti, M., Manso Sainz, R., Landi 
Degl'Innocenti, E., \& Landolfi, M..~2014, \apj, 791, 94 (Paper I)

\bibitem[\protect\citeauthoryear{Gautschi \& Waldvogel}{2001}]{GW01}
Gautschi, W., \& Waldvogel, J.~2001, Computing the Hilbert Transform 
of the Generalized Laguerre and Hermite Weight Functions,
BIT, 41, 490

\bibitem[\protect\citeauthoryear{Hubeny}{1982}]{Hu82}
Hubeny, I.~1982, JQSRT, 27, 593

\bibitem[\protect\citeauthoryear{Landi Degl'Innocenti 
\& Landolfi}{2004}]{LL04}
Landi Degl'Innocenti, E., \& Landolfi, M. 2004, Polarization in
Spectral Lines (Dordrecht: Springer)

\bibitem[\protect\citeauthoryear{Lee}{1977}]{Le77}
Lee, J.-S.~1977, \apj, 218, 857

\bibitem[\protect\citeauthoryear{McKenna \& Nelson}{1986}]{MN86}
McKenna, S.~J., and Nelson, W.~1986, Ap{\&}SS, 125, 103


\bibitem[\protect\citeauthoryear{Mihalas}{1978}]{Mi78}
Mihalas, D. 1978, Stellar Atmospheres, 2nd ed. (San Francisco: Freeman)

\end{thebibliography}
\end{document}